\documentclass{iau} 
\usepackage{graphicx}
\usepackage{rotating}
\usepackage[round]{natbib}
\usepackage[squaren, Gray, cdot]{SIunits}

\title[JD 11.~~Chemical Evolution of Turbulent Multiphase Molecular Clouds] %% give here short title %%
{Chemical Evolution of Turbulent Multiphase \\ Molecular Clouds}

\author[Valeska Valdivia et al.]   %% give here short author list %%
{Valeska Valdivia$^1$,
Patrick Hennebelle$^1$,
Benjamin Godard$^2$,
Maryvonne Gerin$^2$,
Pierre Lesaffre$^2$
\and Jacques Le Bourlot$^2$}

\affiliation{$^1$Universit\'e Paris Diderot, AIM, Sorbonne Paris Cit\'e, CEA, CNRS, F-91191 Gif-sur-Yvette, France. \\ email: {\tt valeska.valdivia@cea.fr} \\[\affilskip]
$^2$Laboratoire de radioastronomie, LERMA, Observatoire de Paris, \'Ecole Normale Sup\'erieure (UMR 8112 CNRS), 24 rue Lhomond, 75231 Paris Cedex 05, France}

\pubyear{2017}
\volume{332}  %% insert here IAU Symposium No.
\jname{Astrochemistry VII -- Through the Cosmos from Galaxies to Planets}
\editors{M. Cunningham, T. Millar \& and Y. Aikawa, eds..}
\begin{document}

\maketitle

\begin{abstract}
Molecular clouds are essentially made up of atomic and molecular hydrogen, which in spite of being the simplest molecule in the ISM plays a key role in the chemical evolution of molecular clouds. Since its formation time is very long, the H$_2$ molecules can be transported by the turbulent motions within the cloud toward low density and warm regions, where its enhanced abundance can boost the abundances of molecules with high endothermicities.

We present high resolution simulations where we include the evolution of the molecular gas under the effect of the dynamics, and we analyze its impact on the abundance of CH$^+$.
\end{abstract}

\firstsection % if your document starts with a section,
\section{Introduction}

Turbulent molecular clouds present clumpy and filamentary structures, where warm and cold phases are interwoven in a very complex structure. The turbulent motions within molecular clouds can accelerate the conversion of atomic to molecular hydrogen in transient over-densities, and subsequently carry these molecules towards diffuse and warm regions \citep{valdivia2016}, where they can participate in endothermic reactions, such as the formation of CH$^+$, which requires molecular hydrogen to be formed efficiently. In  \citet{valdivia2017} we have shown that the warm H$_2$ can enhance the CH$^+$ column densities by a factor 3 to 10.  \\  
Here we present a summary of these two works, as well as a study of the influence of the shielding by dust and the H$_2$ self-shielding on the evolution of the abundance and distribution of molecular hydrogen, as well as the influence of the magnetic field on the structure and distribution of H$_2$. For this study we used a modified version of the \textsc{Ramses} code \citep{teyssier2002} that includes the formation and destruction processes for H$_2$, its thermal feedback, the shielding by dust for the ultraviolet radiation, and the self-shielding due to other H$_2$ molecules. The implementation and the setup are fully described in \citet{valdivia2016}. For the shielding we have used our tree-based method, detailed in \citet{valdivia2014}. \\

\section{Results}
To understand the H$_2$ molecule formation process under the dynamical influence of a highly inhomogeneous structure, we have performed hydrodynamics and magneto-hydrodynamics (MHD) simulations of realistic molecular clouds formed through colliding streams of warm atomic gas in a 50 pc cubic box. 
\subsection{General results}
\begin{figure*}[t]
\centering
%\hspace*{-1.in}   %images are too wide, I reduce margin in 1 inch
  \begin{tabular}{@{}ccc@{}}

5 Myr & 10 Myr & 15 Myr \\
   \includegraphics[trim={1.5cm 0.2cm 5cm 1.5cm},clip,  height=.3\textwidth]{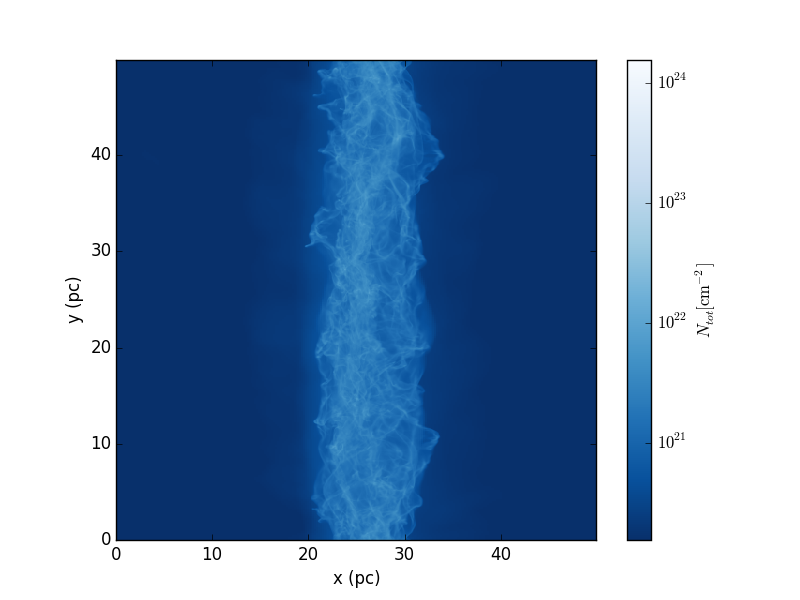} &
   \includegraphics[trim={1.5cm 0.2cm 5cm 1.5cm},clip, height=.3\textwidth]{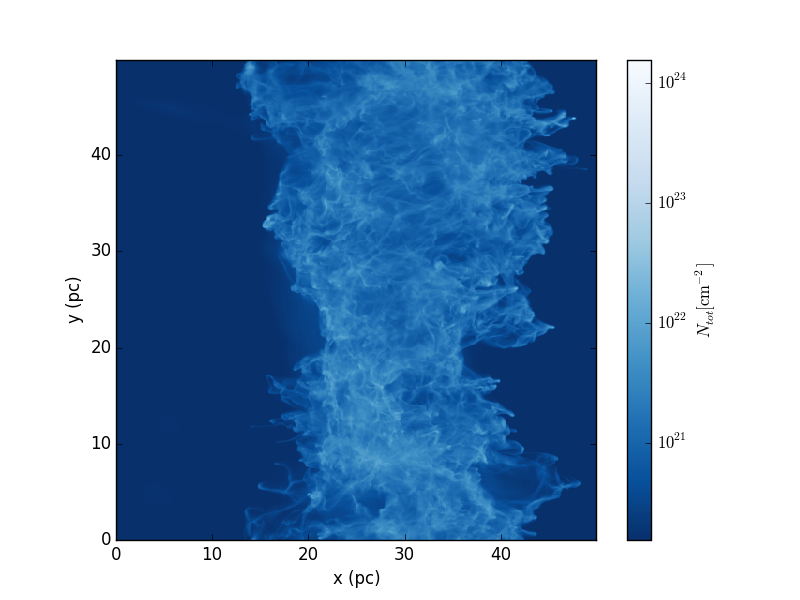} &
   \includegraphics[trim={1.5cm 0.2cm 2.cm 1.5cm},clip,  height=.3\textwidth]{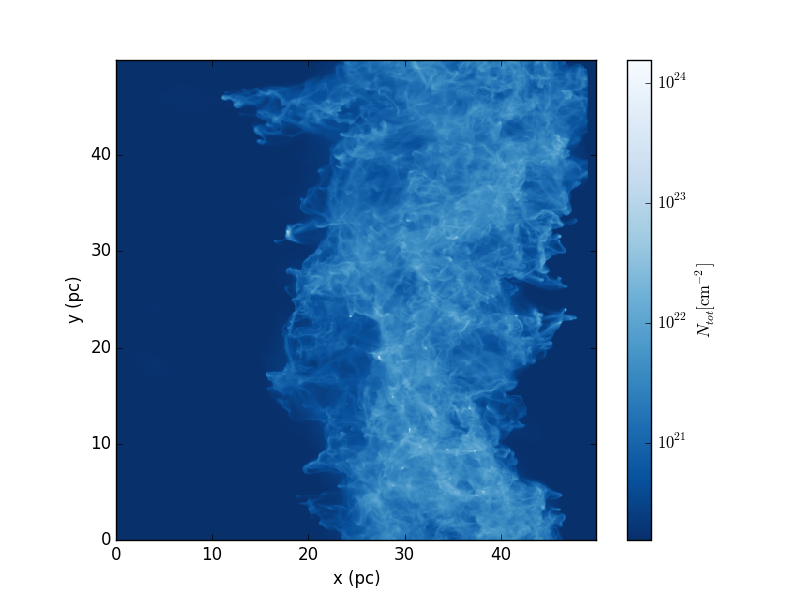} \\
% \begin{sideways} \hspace*{1.5cm} $\mathcal{N}_\mathrm{H_2}$ \end{sideways}&
\includegraphics[trim={1.5cm 0.2cm 5cm 1.5cm},clip, height=.3\textwidth]{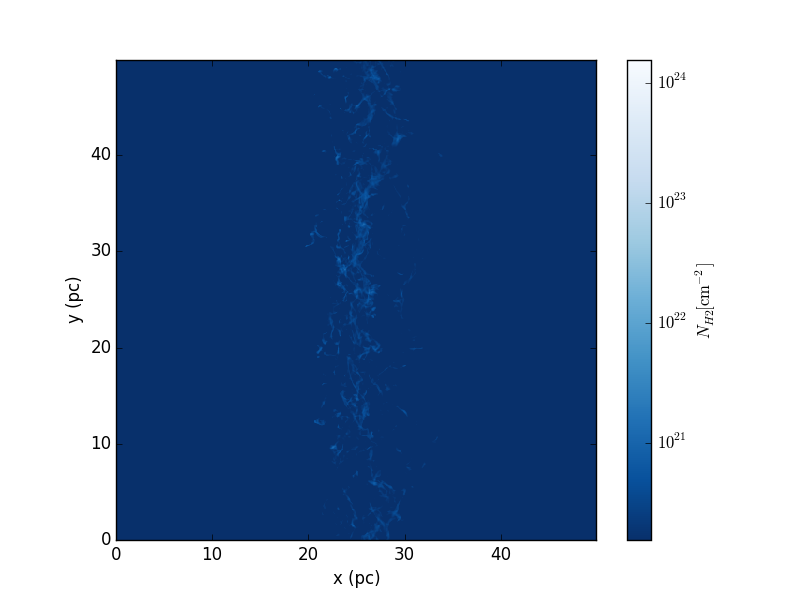} &
\includegraphics[trim={1.5cm 0.2cm 5cm 1.5cm},clip, height=.3\textwidth]{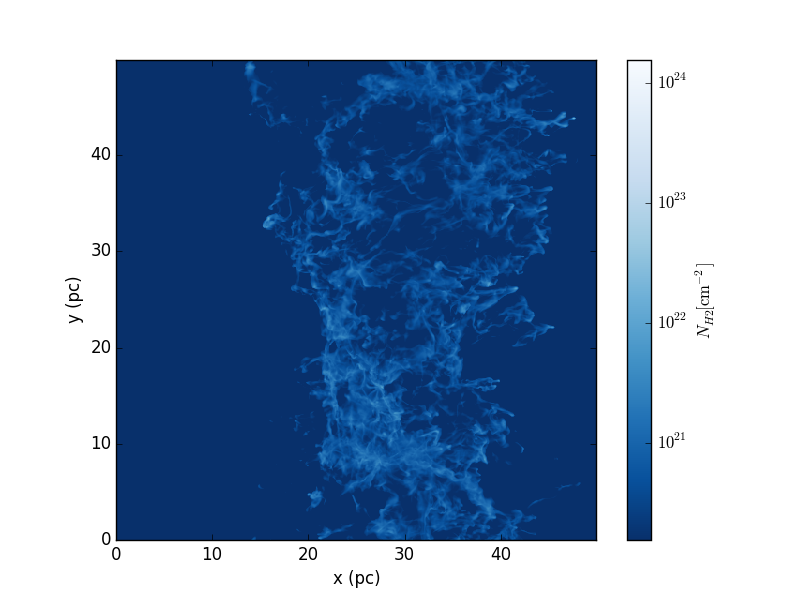} &
\includegraphics[trim={1.5cm 0.2cm 2.cm 1.5cm},clip,  height=.3\textwidth]{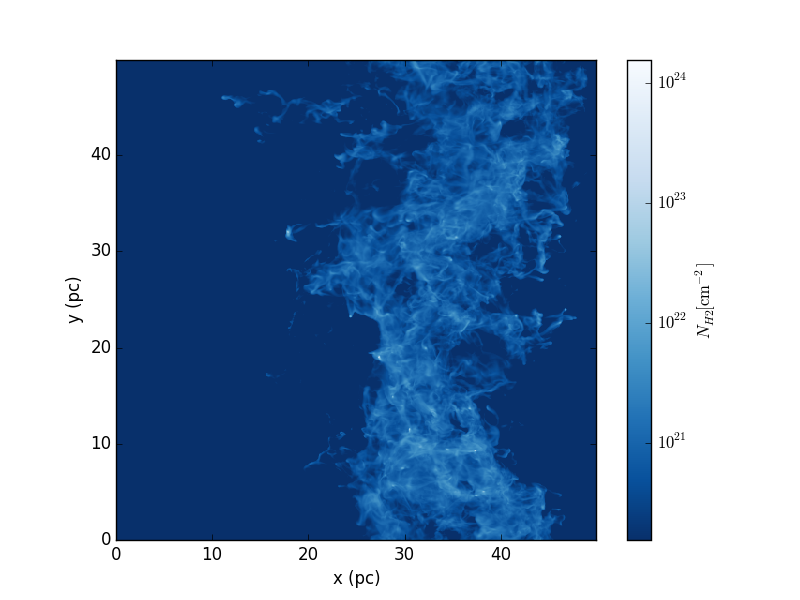} \\
%\begin{sideways} \hspace*{1.5cm} $f(\mathrm{H_2})$ \end{sideways}&
\includegraphics[trim={1.5cm 0.2cm 5cm 1.5cm},clip, height=.3\textwidth]{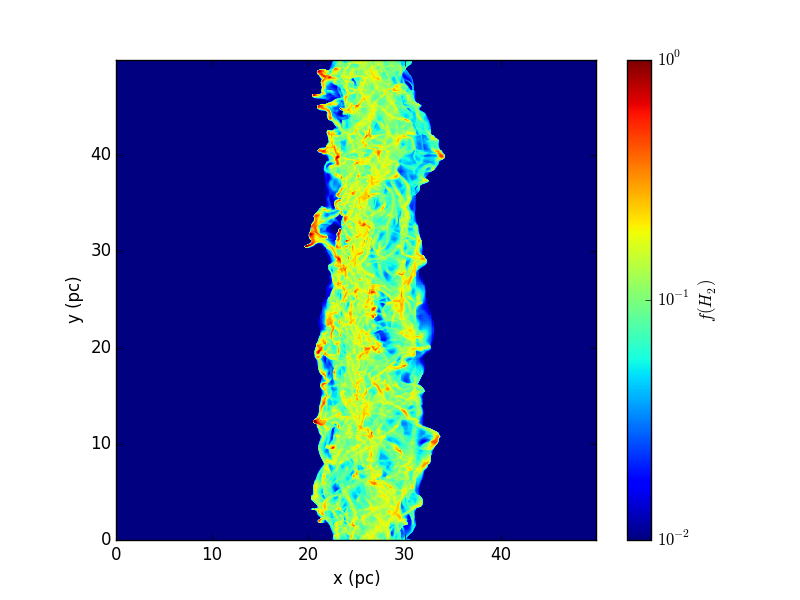} &
\includegraphics[trim={1.5cm 0.2cm 5cm 1.5cm},clip, height=.3\textwidth]{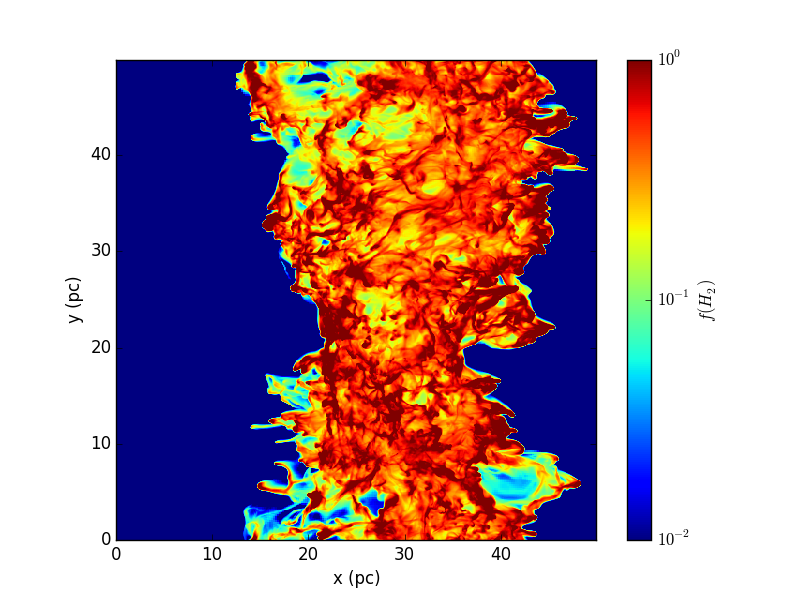} &
\includegraphics[trim={1.5cm 0.2cm 2.cm 1.5cm},clip,  height=.3\textwidth]{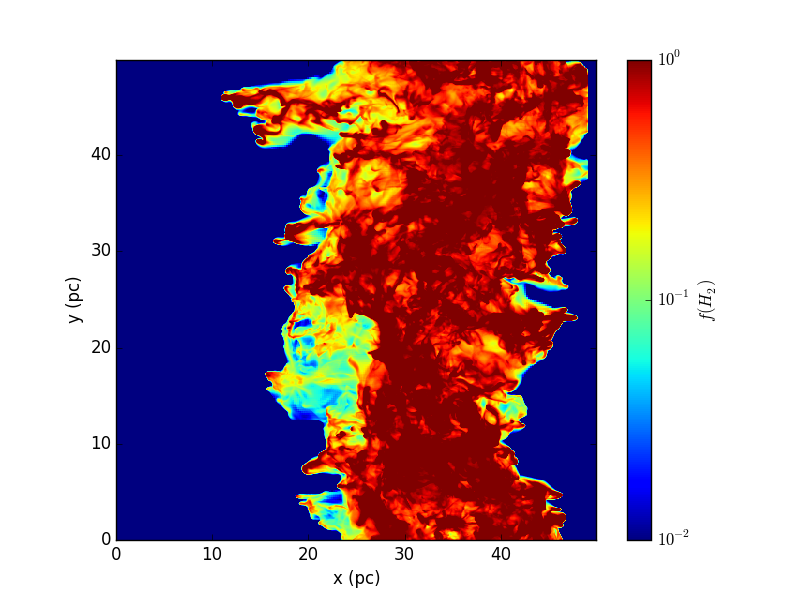} \\
\end{tabular}
\caption{Evolution of a molecular cloud. From top to bottom: Total column density, H$_2$ column density, and the H$_2$ fraction at $t=$ $5$, $10$, and $15$ $\mathrm{Myr}$ (from left to right).}
\label{poster_fig}
\end{figure*}

In \citet{valdivia2016} we have shown that H$_2$ molecular gas is formed faster than the usual estimates based on the mean density of the cloud, which can be interpreted as the result of the local density enhancements induced by turbulent compressions, causing a faster H$_2$ formation \citep{glover2007a, glover2007b, clark2012, micic2012}. Additionally, the relaxation of such over-densities along with the turbulent motions lead to a mixing between different phases and  brings H$_2$ molecules to the warm phase, where they survive due to the shielding provided by the global structure, explaining the presence of warm H$_2$ in the conditions that prevail in molecular clouds. Figure \ref{poster_fig} shows the general evolution of the molecular cloud, as well as the evolution of the molecular hydrogen.\\

\subsection{Influence of the total shielding}

\begin{figure*}[t]
\centering
\hspace*{-0.5cm}
\includegraphics[trim={0.1cm 1.5cm 0.1cm 0.1cm},clip, width=1.05\textwidth]{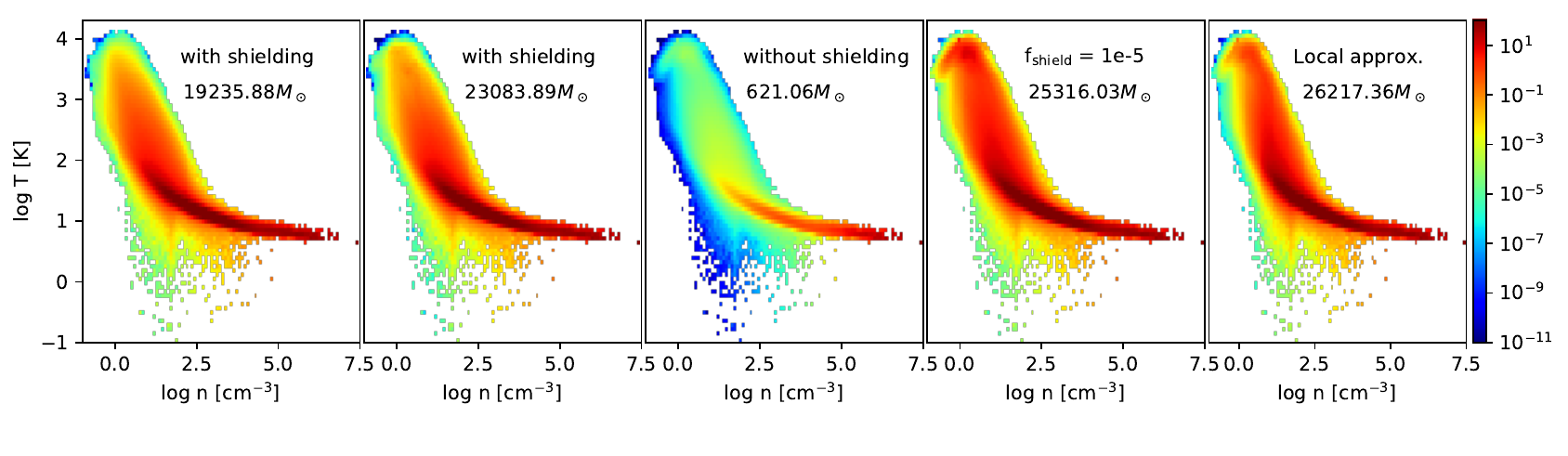}
\caption{Comparison of the normalized $m(\mathrm{H_2}$)-weighted 2D probability density functions (PDF).The leftmost panel shows the result for the simulation including the total shielding. The rest of the panels show the 2D-PDFs expected at equilibrium using different approximations for the total shielding. From left to right: shielding from the actual H$_2$ distribution in the simulation, without any shielding effect, using a constant shielding, and using the local approximation assuming 5\% of H$_2$.}
\label{EqAbund}
\end{figure*}

Nowadays it is widely accepted that molecular hydrogen is formed on dust grain surfaces, and that it is mainly destroyed by UV fluorescent photo-dissociation. The strength of the UV field is attenuated by the presence of dust, and since UV photons are most likely absorbed by other H$_2$ molecules, both the dust and other H$_2$ molecules shield the interior of the molecular cloud against photodissociation. The photo-dissociation rate can then be expressed as : $k_\mathrm{ph} = f_\mathrm{sh, H_2} k_\mathrm{ph,0}$, where $k_\mathrm{ph,0}$ is the unshielded rate \citep{jura1974, gry2002}, and  $f_\mathrm{sh, H_2} = \langle e^{-\tau_{d}}\times f_\mathrm{shield}(\mathcal{N}(\mathrm{H}_2 ))\rangle$ is the total shielding, including dust and self shielding as described in \citet{draine1996}.  Since both shielding mechanisms operate in a line-by-line basis, it is needed to compute them in all directions. This can be extremely expensive in terms of CPU time, so approached methods must be used. To compute total and H$_2$ column densities in our simulation we have used our tree-based method. To test different approximations for the shielding we post-processed a snapshot of our simulation at $15~\mathrm{Myr}$ to calculate the expected abundances of H$_2$ at equilibrium using the actual shielding as well as other less rigorous approximations for the shielding factors. For all cases we used the temperature computed in the simulation. In Fig.~\ref{EqAbund} we compare the H$_2$ mass-weighted 2D probability density functions (PDF). The leftmost panel shows the actual distribution of the dynamically computed H$_2$ from our simulation, while the second panel show the distribution expected at equilibrium for the same shielding conditions. The third panel shows the distribution for the extreme unshielded case. The fourth panel corresponds to the case where the shielding is supposed uniform throughout the whole molecular cloud and equal to the mean shielding factor for gas in the density range between $3$ to $3000~\mathrm{cm}^{-3}$ from Fig.~7 in  \citet{valdivia2016}. The rightmost panel corresponds to the local approximation, where column densities are approximated by $\mathcal{N}\sim n_\mathrm{local}\times L$, where $ n_\mathrm{local}$ is the local total gas number density, and $L$ is the size of the molecular cloud. For the self-shielding by H$_2$ we assume an abundance of 5\%.  The total mass of H$_2$ expected in each case is indicated in the panel in solar masses. This figure suggest that when the shielding effects are not included, only the gas denser than several $100~\mathrm{cm^{-3}}$ is able to form H$_2$ efficiently enough to counterbalance the photodissociation. Another remark is that all the cases that included any treatment of the shielding produced comparable amounts of H$_2$. Nevertheless the treatments at equilibrium tend to overestimate the total abundance of H$_2$.  

\begin{figure}[t]
\centering
  \begin{tabular}{@{}cc@{}}
Shielded & Without shielding\\
   \includegraphics[trim={1.cm 0.2cm 1.4cm 1.cm},clip, width=.45\textwidth]{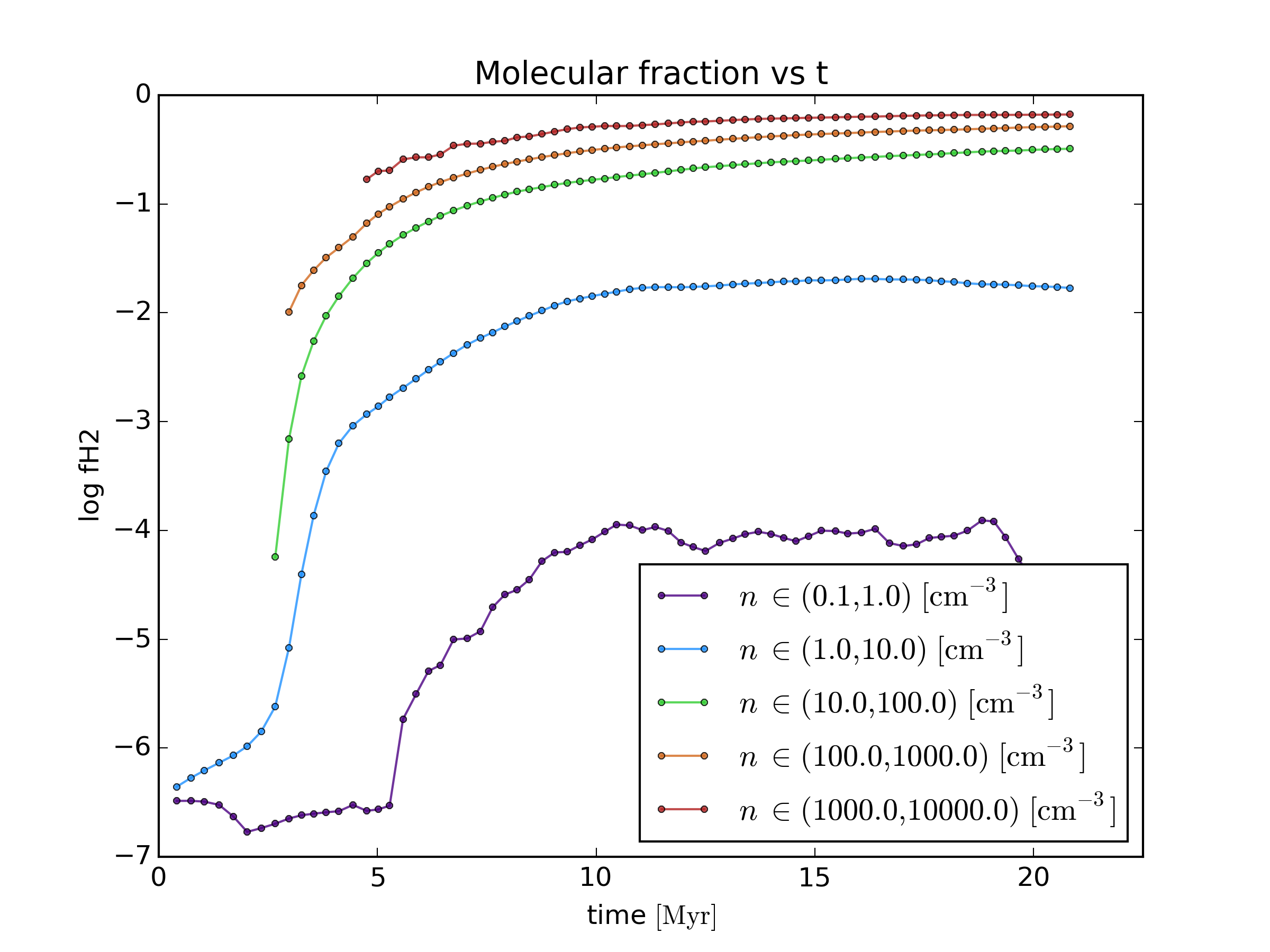} &
   \includegraphics[trim={1.cm 0.2cm 1.4cm 1.cm},clip, width=.45\textwidth]{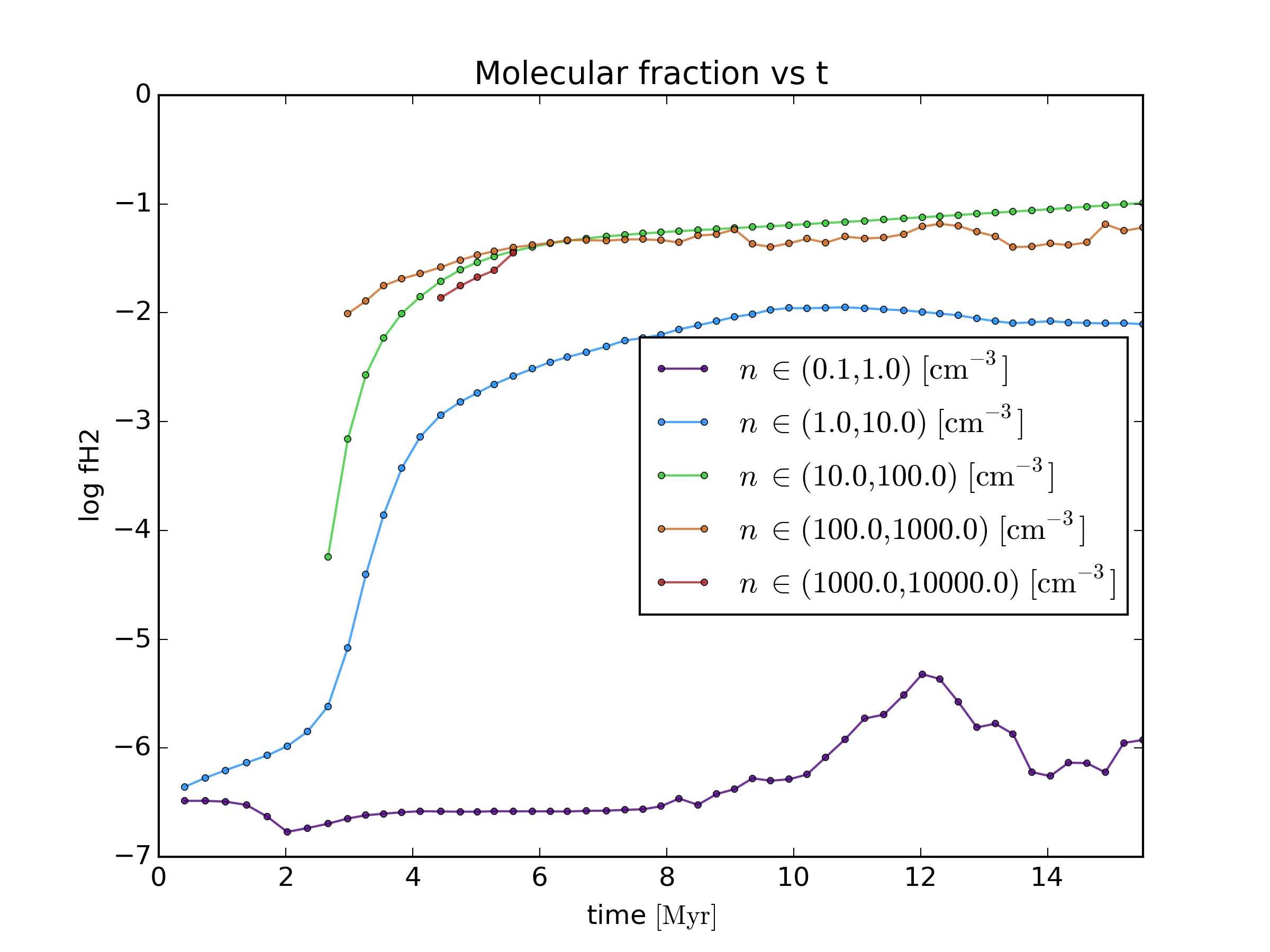} %\\
  \end{tabular}
\caption{Evolution of the fraction of mass in H$_2$ form per density bin: including shielding effects (left) and without shielding (right).}
\label{hydro_EvolutionfH2shield}
\end{figure}

To better understand the influence of the shielding in the dynamical evolution of the cloud we performed two lower resolution simulations ($\sim 0.2~\mathrm{pc}$) using the same setup. Both simulations include the shielding for the UV radiation in the thermal balance, but only the first one includes the total shielding factor $f_\mathrm{sh, H_2} $ for the photodissociation of H$_2$. \\
 Figure~\ref{hydro_EvolutionfH2shield} shows the evolution of the molecular fraction $f(\mathrm{H_2})$ per density bin. 
 When the shielding is not taken into account the presence of molecular gas is globally suppressed, and in a period of $15~\mathrm{Myrs}$ the simulation converts less than $10\%$ of total gas into molecular gas. This is true even for gas denser than several $100~\mathrm{cm^{-3}}$, suggesting that higher densities are not enough to counterbalance the fast destruction when the abundance of H$_2$ is treated dynamically.\\
 We find that when the photodissociation rate includes the combined shielding effects, the amount of gas in H$_2$ form is about $5$ times higher than when this effect is not included. An interesting result comes from the analysis of the most diffuse gas, for which the molecular fraction when the shielding is included is about $2$ orders of magnitude higher than the case without shielding. This indicates that the shielding effect of dust and H$_2$ self-shielding is crucial to explain the presence H$_2$.\\

\subsection{Influence of warm H$_2$ on the abundance of CH$^+$}
Species with long evolution times are most likely influenced by the gas motions, and consequently can be found out-of-equilibrium, whereas species with shorter evolution times can react faster to variations in their environments. Since H$_2$ precedes the formation of other molecules, and since the formation time of H$_2$ is long compared to other molecules, its formation process is a bottleneck for the chemical evolution of the molecular cloud. This permits us to use a hybrid method, fully described in \citet{valdivia2016sf2ab} and \citep{valdivia2017}, where atomic and molecular hydrogen are the only species treated dynamically, while the abundances of the rest of the species are calculated at equilibrium. 
An interesting test case is the presence of CH$^+$ molecules in the diffuse ISM, where the only reaction efficient enough to produce CH$^+$ is highly endothermic \citep[$\Delta E/k = 4300~\mathrm{K}$, ][]{agundez2010}, and involves H$_2$. 

\begin{figure*}[t]
\centering
\hspace*{-0.5cm}
\includegraphics[width=1.05\textwidth]{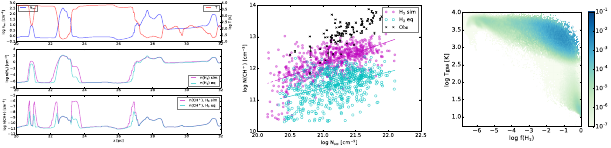}
\caption{Influence of the warm H$_2$. Line-of-sight showing the increased abundance of CH$^+$ at the edge of clumps (left), CH$^+$ column densities (center), and the 2D PDF of CH$^+$ mass as a function of the molecular fraction and the gas temperature (right). The crosses in the central panel are the observational data from \citet{crane1995, gredel1997} and \citet{weselak2008}. \citep[Figures from][]{valdivia2017}}
\label{WarmH2conseq}
\end{figure*}

Figure \ref{WarmH2conseq} shows that the warm H$_2$ have an important influence on the abundance of CH$ ^+$ \citep{valdivia2017, valdivia2016sf2ac}. Most of CH$^+$ is formed in environments where molecular fraction is about a few percent and the gas temperature is higher than $\sim 300~\mathrm{K}$. At the edges of clumps, where H$_2$ is found in excess with respect to the equilibrium value, the column densities of CH$^+$ are higher by a factor of 3 to 10 with respect to the values obtained when H$_2$ is also calculated at equilibrium.

\subsection{Influence of magnetic field}

It is well known that MHD simulations produce clouds with a different morphology than pure hydrodynamical simulations \citep{hennebelleetal2008}. We have performed two simulations using the same set up described in \citet{valdivia2016}, at the same resolution. We use a base resolution of $256^3$ and two additional refinement levels reaching an effective resolution of $\sim 0.05~\mathrm{pc}$. The first one is a pure hydrodynamical simulation ($\vec{B} = 0$), and the second one is a MHD simulation, with a moderate magnetic field aligned with the gas inflow ($B\sim2.5~\mathrm{\micro G}$).  \\

\begin{figure*}[t]
\centering
  \begin{tabular}{@{}ccc@{}}
  5 Myr & 10 Myr & 15 Myr \\
    \includegraphics[trim={1.5cm 0.2cm 5cm 1.5cm},clip, height=.3\textwidth]{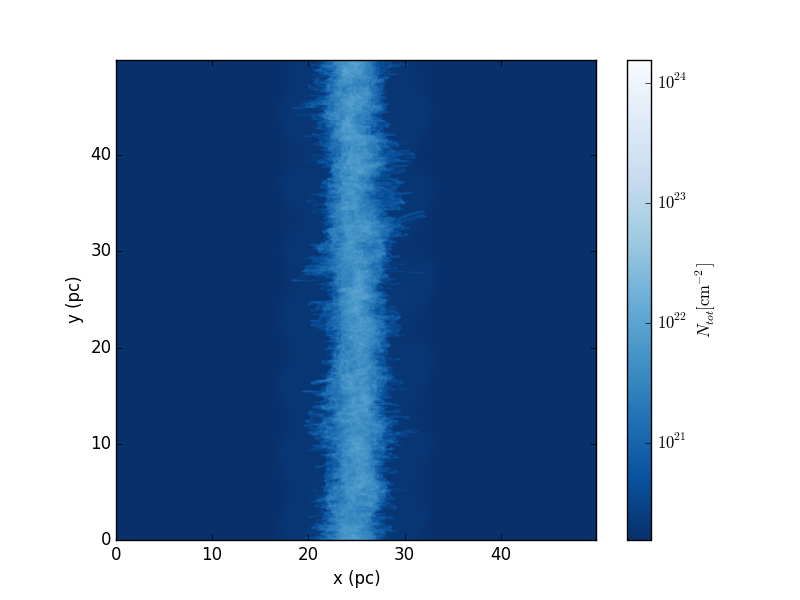} &
        \includegraphics[trim={1.5cm 0.2cm 5cm 1.5cm},clip, height=.3\textwidth]{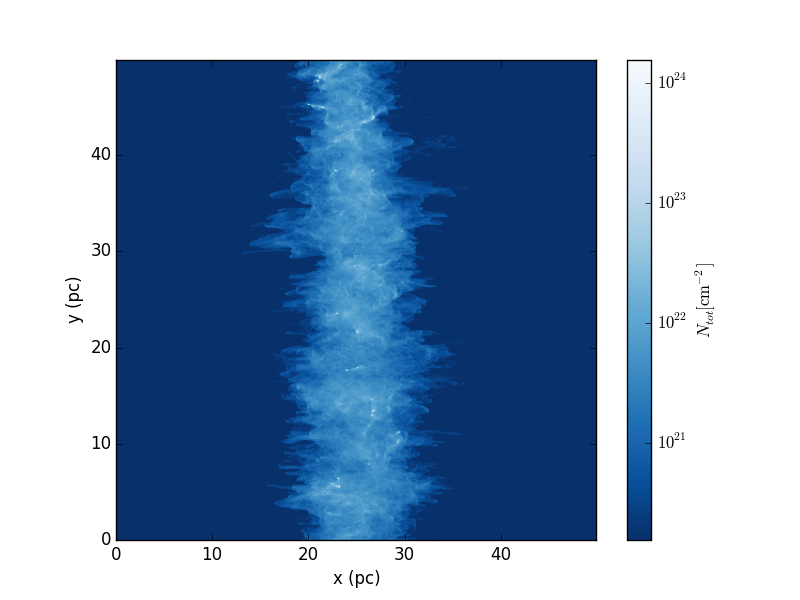} &
\includegraphics[trim={1.5cm 0.2cm 2.cm 1.5cm},clip,  height=.3\textwidth]{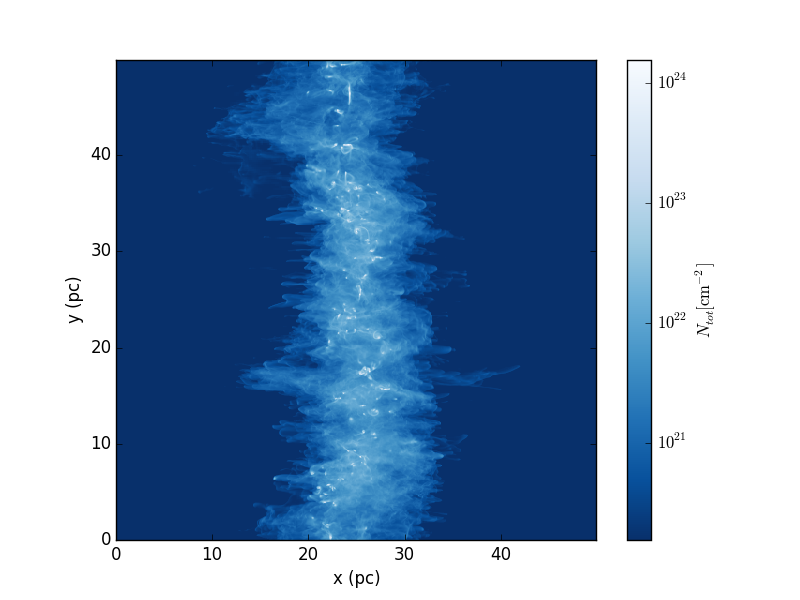} \\    
\includegraphics[trim={1.5cm 0.2cm 5cm 1.5cm},clip, height=.3\textwidth]{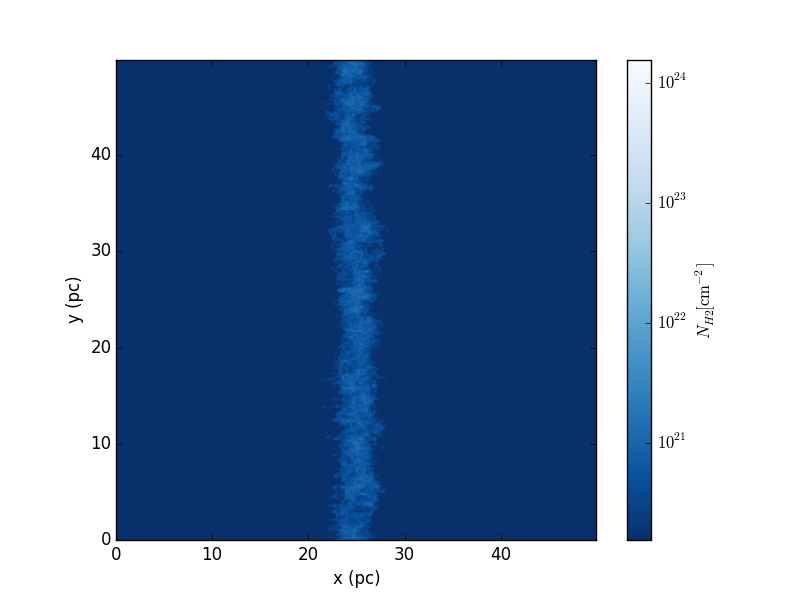} &
\includegraphics[trim={1.5cm 0.2cm 5cm 1.5cm},clip, height=.3\textwidth]{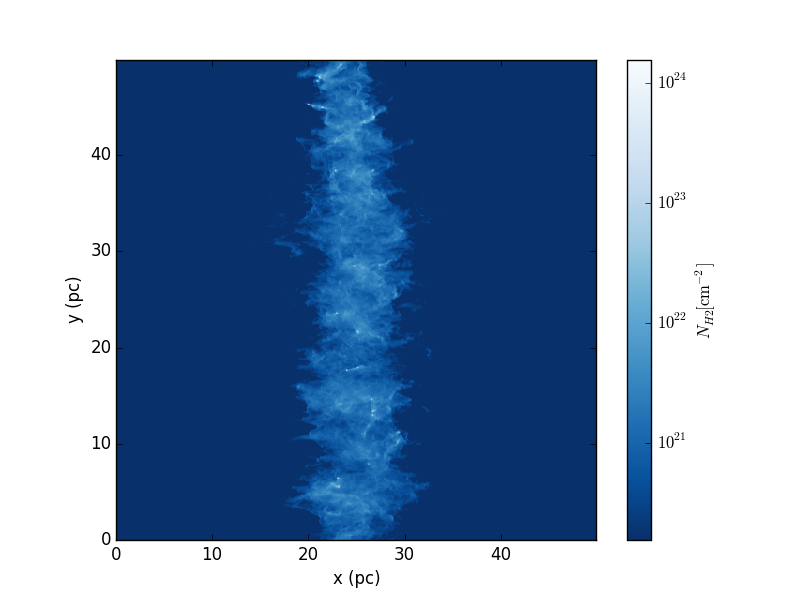} &
\includegraphics[trim={1.5cm 0.2cm 2.cm 1.5cm},clip,  height=.3\textwidth]{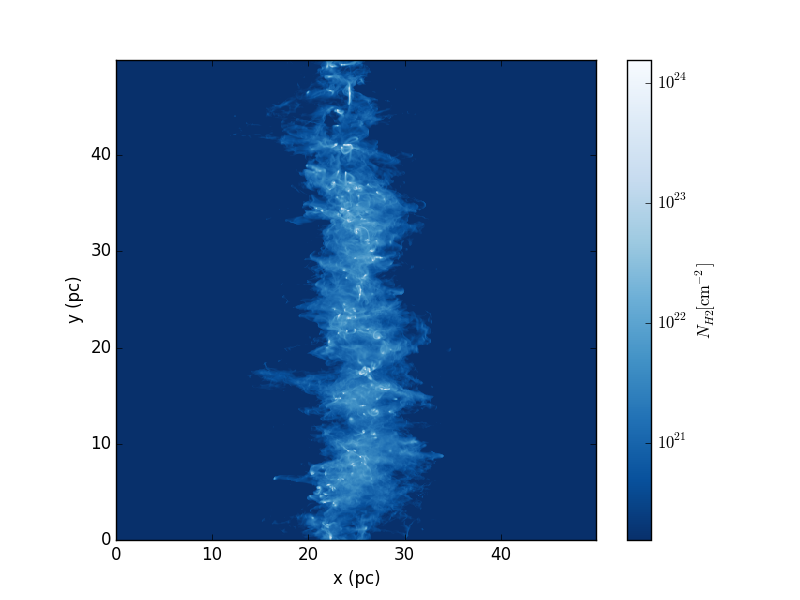} \\
\includegraphics[trim={1.5cm 0.2cm 5cm 1.5cm},clip, height=.3\textwidth]{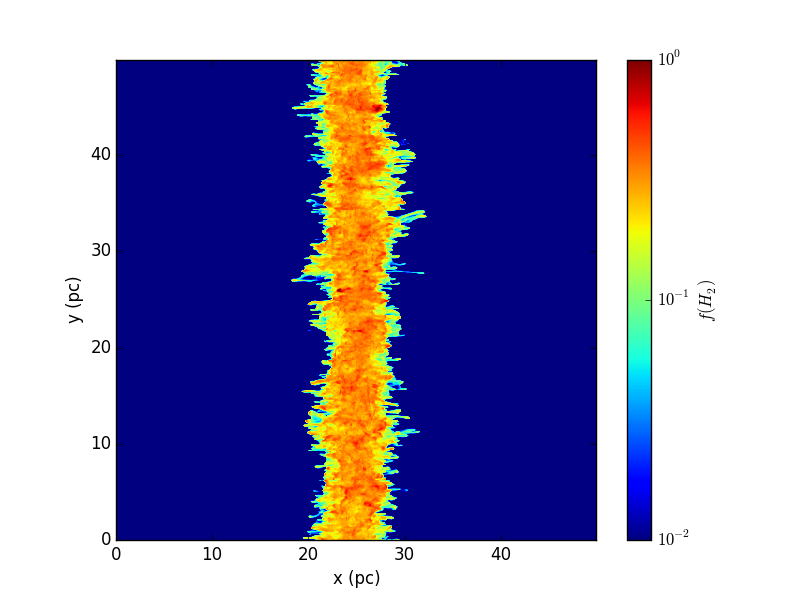} &
\includegraphics[trim={1.5cm 0.2cm 5cm 1.5cm},clip, height=.3\textwidth]{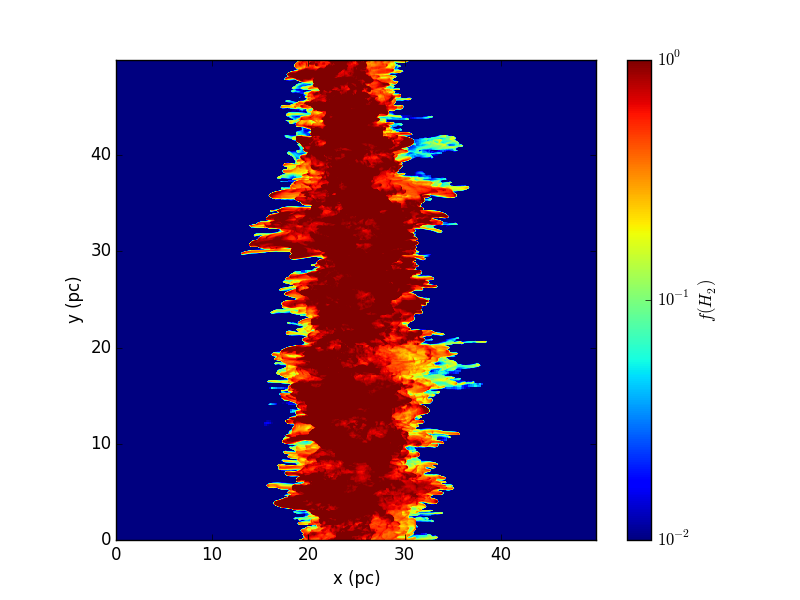} &
\includegraphics[trim={1.5cm 0.2cm 2.cm 1.5cm},clip,  height=.3\textwidth]{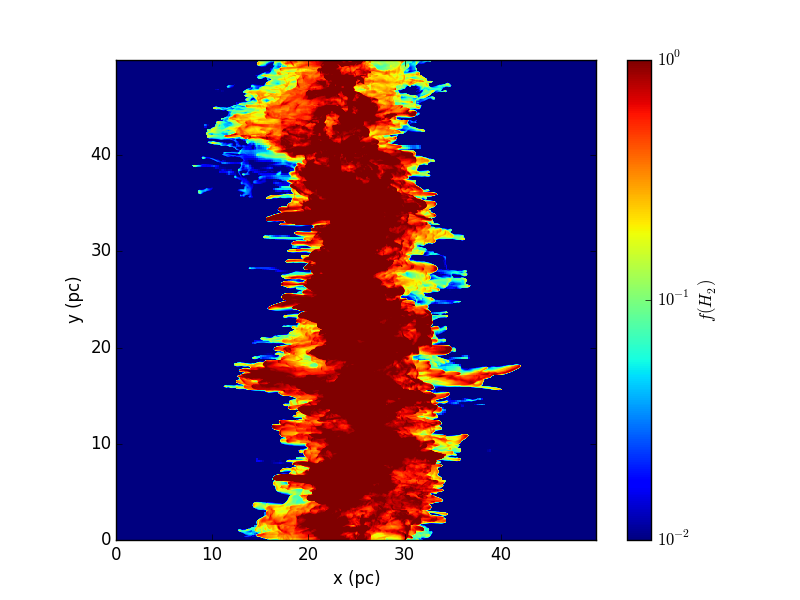} \\
\end{tabular}
\caption{Evolution of a molecular cloud in absence of magnetic field. From top to bottom: Total column density, H$_2$ column density, and the H$_2$ fraction, at $t=$ $5$, $10$, and $15$ $\mathrm{Myr}$.}
\label{hydro_test}
\end{figure*}

\begin{figure}[t]
\centering
  \begin{tabular}{@{}cc@{}}
HYDRO & MHD\\
   \includegraphics[trim={1.cm 0.2cm 1.4cm 1.cm},clip, width=.45\textwidth]{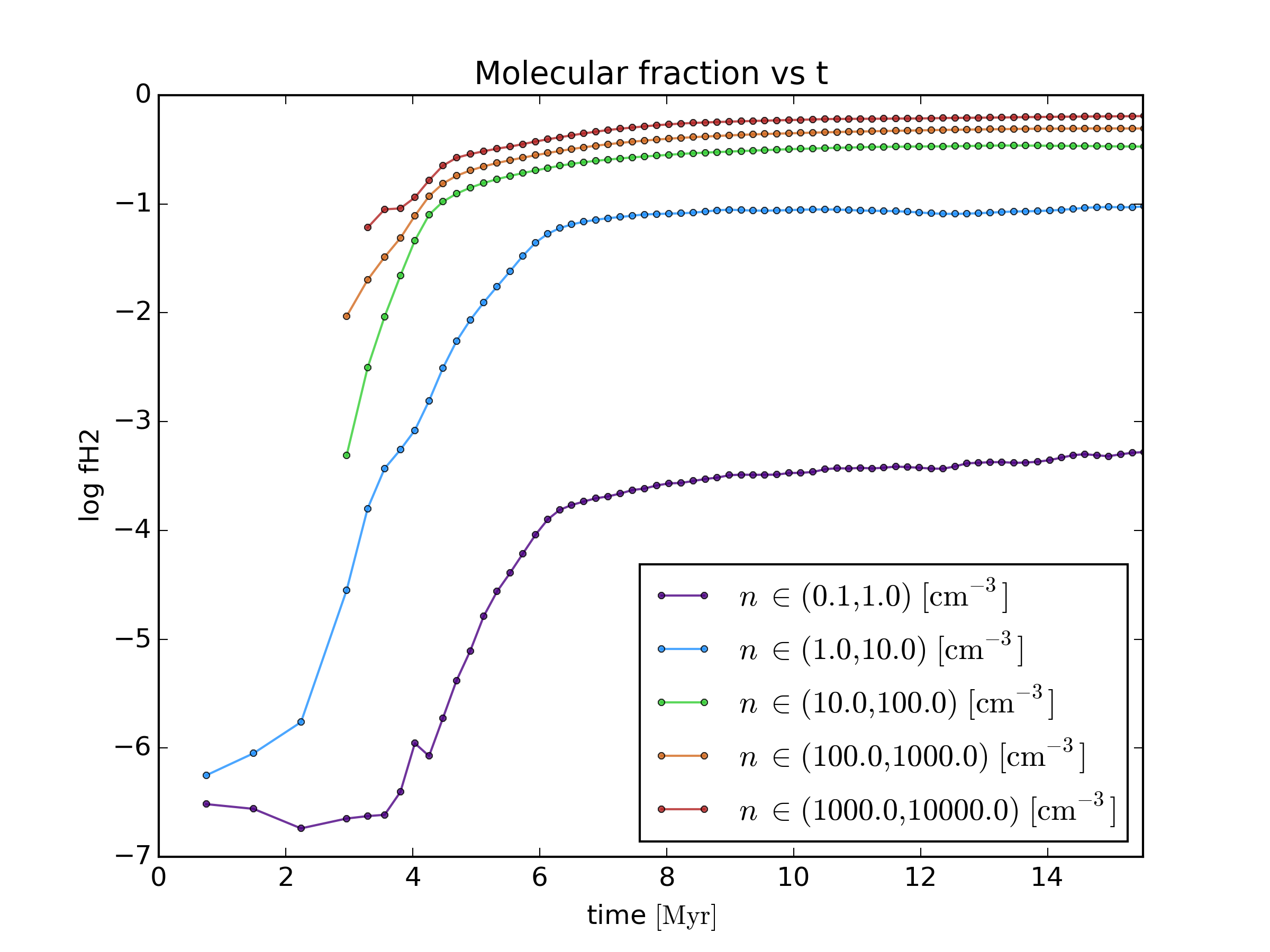} &
   \includegraphics[trim={1.cm 0.2cm 1.4cm 1.cm},clip, width=.45\textwidth]{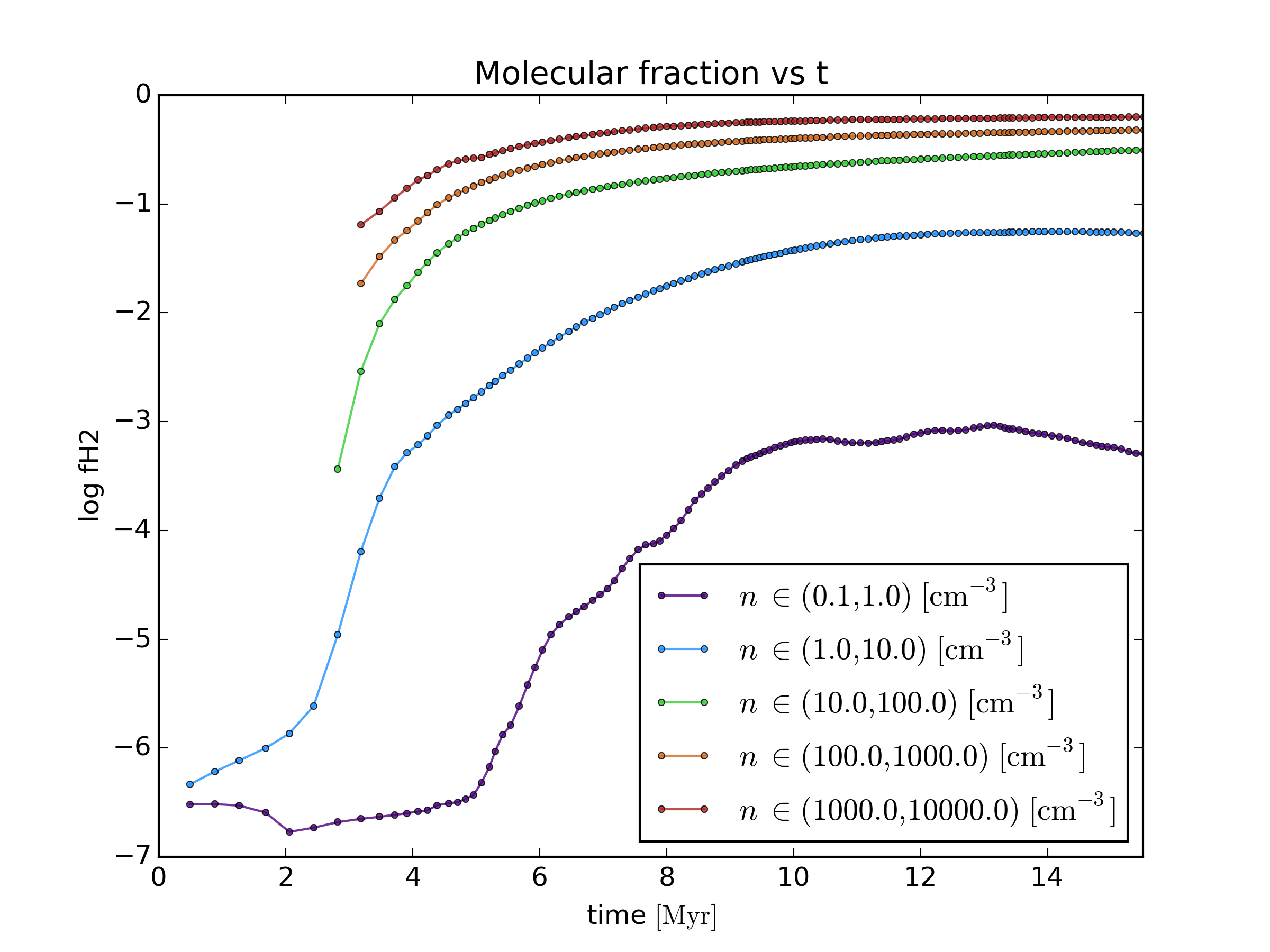} 
  \end{tabular}
\caption{Evolution of the total molecular fraction and the molecular fraction per density bin. Hydro case (left panel), and the standard MHD simulation (right panel).}
\label{hydro_EvolutionfH2}
\end{figure}

Figure \ref{hydro_test} shows the general evolution of the cloud in terms of total column density (top panel) and the H$_2$ colum density (middle panel). The hydro case produces higher column densities at earlier times, as a consequence of the lack of magnetic support. In the MHD simulation, magnetic fields offer an additional support against gravity, delaying the collapse, as shown by \citet{hennebelleetal2008}. The cloud formed in the hydro simulations displays a more compact structure, reaching higher densities that favor an earlier formation of molecular gas (Fig. \ref{hydro_EvolutionfH2}). Higher column densities favor higher molecular fractions, as shown in the bottom panel of Fig. \ref{hydro_test}. An interesting remark is that MHD simulations produce clumpier and more filamentary structures, allowing a more complex mixture of warm and cold gas with a wider range of abundances of H$_2$. \\

Even though the general evolution of the total amount of H$_2$, shown in Fig.~\ref{hydro_EvolutionfH2}, is very similar in both cases, the hydro case reaches $10\%$ of molecular hydrogen about $1~\mathrm{Myr}$ earlier than the MHD case. The difference between both cases is more pronounced at densities of a few $\sim 1 - 10 ~\mathrm{cm^{-3}}$.  \\

\section{Conclusions}
H$_2$ formation can proceed at early stages of the molecular cloud evolution due to compressive motions driven by the turbulence \citep{micic2012}. The turbulent motions within the gas can mix the warm neutral and the cold neutral phases and transport long-lived molecules towards diffuse and warm environments. 

The dust and the surrounding  H$_2$ molecules shield cloud interior against the dissociating UV radiation, protecting the H$_2$ present in diffuse warm gas. The presence of magnetic fields offers a magnetic support against the gravitational collapse, delaying the formation of dense structure, and thus delaying the formation of H$_2$. Nevertheless, magnetized gas produces more complex and more filamentary structures displaying a wide variety of physical conditions.  

The warm H$_2$ is an important ingredient in highly endothermic reactions, increasing the column densities of CH$^+$ by a factor $3$ to $10$.  Locally this effect is even more striking, being about two orders of magnitude more abundant at the edge of clumps where there is an excess of H$_2$ with respect to the value expected at equilibrium. Nevertheless total column densities are still under-predicted by a factor of roughly 6, suggesting than other physical process, such as the dissipation of turbulence, may be at play \citep{godard2014}.

\end{document}